\documentclass[lettersize,journal]{IEEEtran}
\usepackage{amsmath,amsfonts}

\usepackage{cite}
\usepackage{array}

\usepackage{tabularx}
\usepackage{booktabs}

\usepackage{bbding}
\usepackage[caption=false,font=normalsize,labelfont=sf,textfont=sf]{subfig}
\usepackage{multirow}
\usepackage{makecell}
\usepackage{textcomp}
\usepackage{stfloats}
\usepackage{url}
\usepackage{verbatim}
\usepackage{graphicx}
\usepackage[T1]{fontenc} 
\usepackage{amssymb} 
\usepackage{algorithmic}
\usepackage{algorithm}
\usepackage[backref=False]{hyperref}
\def\BibTeX{{\rm B\kern-.05em{\sc i\kern-.025em b}\kern-.08em
    T\kern-.1667em\lower.7ex\hbox{E}\kern-.125emX}}
\usepackage{balance}
\newcommand{\Rmnum}[1]{\uppercase\expandafter{\romannumeral #1}}  
\usepackage{tikz,xcolor,hyperref}

\usepackage{amsmath}
\usepackage{subcaption}
\usepackage{graphicx}
\usepackage{amssymb}

\definecolor{lime}{HTML}{A6CE39}
\DeclareRobustCommand{\orcidicon}{
	\begin{tikzpicture}
		\draw[lime, fill=lime] (0,0)
		circle[radius=0.16]
		node[white]{{\fontfamily{qag}\selectfont \tiny \.{I}D}}; 
	\end{tikzpicture}
	\hspace{-2mm}
}
\foreach \x in {A, ..., Z}{%
	\expandafter\xdef\csname orcid\x\endcsname{\noexpand\href{https://orcid.org/\csname orcidauthor\x\endcsname}{\noexpand\orcidicon}}
}

\begin{document}
\title{Hierarchical Optimization Based Multi-objective Dynamic Regulation Scheme for VANET Topology}

\author{Ruixing~Ren\hspace{-1.5mm}\orcidA{}, Minqi~Tao\hspace{-1.5mm}\orcidB{}, Junhui~Zhao\hspace{-1.5mm}\orcidC{},~\IEEEmembership{Senior~Member,~IEEE}, Xiaoke~Sun\hspace{-1.5mm}\orcidD{}, and Qiuping~Li\hspace{-1.5mm}\orcidE{}

\thanks{Corresponding author: Junhui Zhao.

Ruixing Ren, Junhui Zhao are with the School of Electronic and Information Engineering, Beijing Jiaotong University, Beijing 100044, China. (e-mail: renruixing0604@163.com)

Minqi Tao is with the School of Information and Software Engineering, East China Jiaotong University, Nanchang 330013, Chiina.
		
Xiaoke Sun and Qiuping Li are with the National Computer Network Emergency Response Technical Team/Coordination Center of China (CNCERT/CC), Beijing 100029, China.}
}

\maketitle

\begin{abstract}
  As a core technology of intelligent transportation systems, vehicular ad-hoc networks support latency-sensitive services such as safety warning and cooperative perception via vehicle-to-everything communications. However, their highly dynamic topology increases average path length, raises latency, and reduces throughput, severely limiting communication performance. Existing topology optimization methods lack capabilities in multi-objective coordination, dynamic adaptation, and global-local synergy. To address this, this paper proposes a two-layer dynamic topology regulation scheme combining local feature aggregation and global adjustment. The scheme constructs a dynamic multi-objective optimization model integrating average path length, end-to-end latency, and network throughput, and achieves multi-index coordination via link adaptability metrics and a dynamic normalization mechanism. it quickly responds to local link changes via feature fusion of local node feature extraction and dynamic neighborhood sensing, and balances optimization accuracy and real-time performance using a dual-mode adaptive solving strategy for global topology adjustment. It reduces network oscillation risks by introducing a performance improvement threshold and a topology validity verification mechanism. Simulation results on real urban road networks via the SUMO platform show that the proposed scheme outperforms traditional methods in average path length (stabilizing at ~4 hops), end-to-end latency (remaining ~0.01 s), and network throughput.
\end{abstract}

\begin{IEEEkeywords}
Vehicular ad-hoc network; network governance; topology optimization; multi-objective optimization
\end{IEEEkeywords}

\section{Introduction}
As a core technology of intelligent transportation systems, vehicular ad-hoc networks (VANETs) provide critical support for services including safety warning, emergency braking and cooperative perception via vehicle-to-vehicle (V2V), vehicle-to-infrastructure (V2I) and vehicle-to-cloud \cite{RenITS,VANETSurvey}. These applications are highly latency-sensitive, requiring reliable millisecond-scale information transmission \cite{RenIoV}.

Average path length, a core metric for network transmission efficiency, denotes the mean number of hops required for communication between any two nodes \cite{Tao,10483853}. In static networks, shorter paths correspond to lower latency and higher reliability. However, VANETs exhibit high dynamics: high-speed vehicle movement, complex road environments, and traffic flow fluctuations collectively drive continuous, drastic topology changes \cite{RenUAV}, causing frequent disruptions to optimal communication links. Packets are forced to reroute or add forwarding hops, triggering a surge in average path length \cite{Hussein}. This not only directly increases end-to-end delay and intensifies resource competition, but also risks network congestion, severely limiting VANETs’ practical service capabilities \cite{Identity}.

To address the above challenges, existing studies have developed multiple technical approaches for topology optimization. Traditional optimization algorithms, with mathematical modeling as their core, offer computational simplicity advantages in simple scenarios. Greedy algorithms rely on local optimal decisions; although they can rapidly reduce link conflicts, they tend to fall into local optima and fail to balance path length and stability in high-density vehicle scenarios \cite{Greedy_1,Greedy_2}. Integer programming can derive theoretical global optimal solutions via modeling, yet its computational complexity grows exponentially with node count, failing to meet real-time requirements \cite{IP_1,IP_2}. Graph theory-based shortest path methods (e.g., Dijkstra, A*) aim to minimize hop count, but often overlook frequent link disruptions caused by high-speed vehicle movement \cite{Graph_1,Graph_2}.

Intelligent optimization algorithms enhance topology adaptability in dynamic scenarios by simulating natural mechanisms or implementing autonomous learning. Nature-inspired algorithms (e.g., African vulture optimization \cite{AVOCA}, snow goose, pigeon-inspired optimization) simulate swarm intelligence to optimize topology parameters, performing well in complex problems, yet they generally converge slowly and cannot promptly respond to traffic flow mutations. Multi-agent reinforcement learning adopted by Sun et al.\cite{MARL} adapts to large-scale dynamic networks via distributed decision-making, and mitigates computational pressure using the centralized Critic-distributed Actor architecture. However, it remains constrained by low decision-making efficiency in high-dimensional state spaces and insufficient agent collaboration, failing to simultaneously meet VANET’s real-time and global optimization requirements.

Distributed clustering strategies partition networks into relatively stable cluster structures, aiming to mitigate the impact of topological dynamics. For cluster head election, the work in \cite{AVOCA} optimizes cluster size to reduce overhead. For cluster maintenance, the distributed cluster topology maintenance scheme proposed by Gäbler et al.\cite{Cluster_1} enhances network continuity by dynamically adjusting structures and services. The method proposed for UAV ad-hoc networks\cite{Cluster_2} further integrates a reinforcement learning module to dynamically optimize clustering strategies based on node utility, and uses inter-cluster forwarding nodes to improve transmission performance, optimizing both clustering efficiency and quality of service. However, clustering strategies still have notable limitations: frequent cluster head switching increases signaling overhead, structural adjustments lag behind traffic mutations, and their adaptability is generally insufficient in heterogeneous scenarios (e.g., high-speed movement, dense congestion).

For the high dynamics of VANET topologies, specialized regulation methods enhance adaptability primarily via active intervention or dynamic modeling. The centralized periodic intervention networking method proposed by Wu et al. \cite{Wu} performs routing optimization with periodic intervention as the core framework; it adapts to dynamic scenarios via long-time-domain mapping and active optimization mechanisms, balancing routing real-time performance and robustness over long time domains. Dynamic topology evolution modeling methods can evaluate routing stability by capturing continuous network topology changes. For example, the dynamic topology evolution model proposed by Wang et al. \cite{Wang} integrates this evolution model with routing optimization, providing predictive support for routing decisions based on topology changes. However, these methods do not explicitly incorporate the synergy mechanism between local link changes and global topology optimization in their design, leaving room for improvement in balancing multi-objective performance. Additionally, the adaptive VANET topology optimization method proposed by Xiao et al.\cite{Xiao} partitions logical spaces via adaptive K-means clustering to improve V2V and V2I communication efficiency, though the real-time performance of clustering partitioning in dynamic scenarios still needs enhancement.

Based on above research, current VANET topology optimization methods still have three common gaps. First, insufficient multi-objective collaborative optimization: most schemes focus on a single metric (e.g., path length, cluster stability) and overlook the comprehensive balance of latency, throughput, and resource consumption \cite{Wang,Wu}. Second, limited dynamic adaptability: traditional algorithms struggle to cope with rapid topological changes, while intelligent algorithms and clustering strategies suffer from real-time performance or generalization deficiencies \cite{MARL,Cluster_1}. Third, lack of local-global synergy mechanisms: local adjustments tend to fall into suboptimal solutions, while global optimization faces conflicts between complexity and real-time performance \cite{Xiao,Wu}. 

To address these gaps, this paper proposes a two-layer dynamic topology regulation scheme integrating local feature aggregation and global adjustment: the local layer rapidly responds to local link changes caused by node movement to avoid the adjustment lag of traditional algorithms; the global layer coordinates multi-objective optimization goals to break through the performance bottlenecks of local optimization methods such as greedy algorithms. This scheme not only absorbs the local stability advantages of distributed clustering strategies \cite{AVOCA,Cluster_1}, but also draws on the global optimization ideas of intelligent optimization algorithms \cite{MARL}. Meanwhile, it balances computational complexity and real-time performance through a hierarchical architecture, providing a new solution for efficient communication in VANET high-dynamic scenarios. The main contributions of this paper are as follows:
\begin{itemize}
	\item A dynamic multi-objective optimization model integrating average path length, end-to-end delay and network throughput is constructed; combined with link adaptability metrics and a dynamic normalization mechanism, it achieves multi-index collaborative optimization.
	\item A hierarchical optimization-based dynamic topology regulation scheme is proposed, which balances optimization accuracy and real-time performance via local node feature extraction, dynamic neighborhood-aware feature fusion, and global topology adjustment. A performance improvement threshold and topology validity verification are introduced to reduce network oscillation risks.
	\item Simulation verification using real road network maps imported via SUMO shows that the proposed scheme outperforms traditional methods in average path length, end-to-end delay and network throughput, providing key technical support for the reliable deployment of VANETs in dynamic traffic scenarios.
\end{itemize}

The remainder of this paper is organized as follows. Section \ref{2} establishes the system model and formalizes the decision variables and objective functions of hierarchical optimization. Section \ref{3} presents the hierarchical optimization-based dynamic VANET topology regulation scheme. Section \ref{4} compares it with other methods via simulation experiments to quantitatively analyze the optimization effect. Section \ref{5} concludes the work.

\section{System Model}\label{2}
This section defines the dynamic network topology and performance metrics of VANETs, formulates a joint optimization problem of end-to-end delay and average path length, and mitigates communication congestion in VANETs.

\subsection{Dynamic VANET Topology}
The dynamic topology of VANETs is denoted as \(\mathcal{G}(t)=(U(t),E(t))\), which consists of the full node set \(U(t)\) and communication link set \(E(t)\) in the network at time $t$. Vehicles are the basic communication entities of VANETs; each is equipped with an on board unit (OBU) that can collect and update its own status information in real time. Due to the high mobility of vehicles, the vehicle nodes within the network communication coverage at time t change dynamically. Thus, the total number of vehicle nodes in the network at time $t$ is denoted as \(N(t)\), and the set of mobile vehicle nodes is expressed as $\mathcal{V}(t)=\left \{ v_1(t),...,v_n(t),...,v_N(t) \right \} $.

Road side units (RSUs) are fixed infrastructure nodes in VANETs, with their deployment locations and quantities remaining stable throughout the network operation period without temporal dynamic changes. Denoting the total number of RSUs in the network as 
$M$, the RSU node set is expressed as $\mathcal{R} =\left \{ r_1,...,r_m,...,r_M \right \} $. Deployed in segments along road sections, each RSU manages the communication and data of vehicles within its coverage area, and can collect status data (e.g., position, speed, communication demands) reported by OBUs via dedicated short-range communication or cellular vehicle-to-everything interfaces. Meanwhile, RSUs are interconnected via high-speed networks, enabling real-time collaboration for cross-region data fusion and joint decision-making with negligible interaction and processing delay. Based on the above definitions, the full node set of the VANET at time $t$ can be expressed as $U(t)=\mathcal{V}(t)\cup\mathcal{R}$. 

To select V2V communication links with high stability, a link adaptability index $R_{n,u}$ between vehicles $v_n(t)$ and $v_u(t)$ ($u\in \mathcal{V}$) is defined, which is determined by the similarity of the vehicles' motion states (the time index $t$ is omitted):
\begin{equation}
	R_{n,u} = \alpha \cdot \frac{\min(s_n,s_u)}{max(s_n,s_u)}+(1-\alpha)\cdot \cos|\theta_n-\theta_u|
\end{equation}
where \(\alpha \in [0,1]\) denotes the velocity similarity weight coefficient, which balances the impacts of velocity and direction on link stability. $s_n$ and $s_u$ represent the velocities of vehicles $v_n$ and $v_u$, respectively. $\theta_n$ and $\theta_u$ are the driving direction angles of the two vehicles. When \(R_{n,u} \geq R_{\text{th}}\), a V2V communication link is preferentially established between vehicles $v_n$ and $v_u$, where $R_{\text{th}}$ is the trajectory coincidence threshold.

Multiple RSUs execute a distributed collaborative optimization algorithm based on collected and fused global information to jointly generate the dynamic link adjustment strategy $S(t)$ for VANET, which mainly includes link control variables in two dimensions: V2V and V2I. A binary variable $x_{n,u}$ is defined as the V2V link activation variable. If $x_{n,u}=1$ the V2V link between vehicle nodes $v_n(t)$ and $v_u(t)$ is activated, otherwise it is disconnected. A binary variable $y_{n,m}(t)$  is defined as the V2I link activation variable with the same setting as $x_{n,u}(t)$. A bandwidth allocation parameter $b_{n,m}$ for V2I links is introduced to denote the bandwidth resources actively allocated by $r_m$ to vehicle $v_n$. Each RSU node broadcasts $S(t)$ to the OBUs of all vehicles within its coverage area, driving the dynamic reconfiguration of V2V and V2I links across the entire network, and finally forming an optimized VANET topology $G^{\ast}(t)$, where the edge set $E^{\ast}(t)$ represents the adjusted effective communication links.

The communication link set \(E(t)\) of VANET at time $t$ can be further decomposed into the V2V link set \(E_{\text{V2V}}(t)\) and the V2I link set \(E_{\text{V2I}}(t)\), satisfying \(E(t) = E_{\text{V2V}}(t) \cup E_{\text{V2I}}(t)\). In practical VANET communication scenarios, all links are subject to a maximum connection number constraint, which is jointly determined by the hardware communication performance of OBUs and RSUs, as well as limited spectrum resources. Meanwhile, the stability of VANET communication links is synergistically affected by multiple factors including high-speed vehicle movement, limited communication transmission distance, wireless channel interference and environmental noise \cite{Deng}, rendering the link set $E(t)$ with significant time-varying characteristics and ultimately endowing the VANET topology $\mathcal{G}(t)$ with high dynamism.

\subsection{Definition of Performance Metrics}
To evaluate the congestion level of VANET and guide topology optimization, this work selects average path length, end-to-end delay, and network throughput as core metrics to quantify the performance improvement degree after topology adjustment.

\noindent\textit{(1) End-to-End Delay}

For the communication demand from any source vehicle node $v_s(t)$ to destination node $v_d(t)$, its transmission path $P_{sd}(t)$ can be a V2V multi-hop path, a V2I path, or a hybrid path. Accordingly, its end-to-end delay $T_{sd}(t)$ consists of three components: queuing delay, transmission delay, and propagation delay.

Queuing delay refers to the time that data packets wait for scheduling in the buffer queues of each relay node along the transmission path; it is mainly determined by the current load of the node and directly reflects the congestion degree of the relay node \cite{roy2021overview}, which includes two types of relay nodes: vehicles and RSUs. If the data packet is currently at the vehicle relay node \(v_n(t)\) (\(v_n(t) \in P_{sd}(t)\) and $v_n(t) \neq v_d(t)$, i.e., a vehicle relay that is not the destination node), its queuing delay is given by:
\begin{equation}
	T_{\text{queue},n}^V(t) = k_V \cdot \deg_{\text{in}}(v_n(t),t) \cdot \tau^V
\end{equation}
where \(\deg_{\text{in}}(v_n(t),t)\) denotes the number of active links pointing to the vehicle relay node at time $t$ (reflecting the node load), \(\tau^V\) is the average processing time of a single data packet at a vehicle node, and \(k_V\) is the proportionality coefficient corresponding to the scheduling mechanism of the vehicle node.

If the data packet is currently at the RSU relay node $r_m(t) \in P_{sd}(t)$, its queuing delay is given by:
\begin{equation}
	T_{\text{queue},m}^I(t) = k_I \cdot \deg_{\text{in}}(r_m(t),t) \cdot \tau^I
\end{equation}
where \(\deg_{\text{in}}(r_m(t),t)\) denotes the number of active links pointing to the RSU relay node at time $t$, \(\tau^I\) is the average processing time of a single data packet at an RSU, and $k_I$ is the dedicated coefficient corresponding to the RSU scheduling mechanism.

Transmission delay refers to the time taken for a data packet to be transmitted from a sending node (including source nodes and relay nodes) to a receiving node via a relay link, which is related to the available link bandwidth and data packet size \cite{iqbal2023rltd}. For a V2V link \((v_n(t), v_u(t))\) composed of vehicle relay nodes in the path, its transmission delay is given by:
\begin{equation}
	T_{\text{trans},n,u}^V(t) = \frac{L}{B_{n,u}(t)}
\end{equation}
where $L$ is the fixed length of a data packet, and \(B_{n,u}(t)\) is the available bandwidth of the V2V link at time $t$. For a V2I link \((v_n(t), r_m(t))\) composed of a vehicle and an RSU in the path, its transmission delay is given by:
\begin{equation}
	T_{\text{trans},n,m}^I(t) = \frac{L}{B_{n,m}(t)}
\end{equation}
where \(B_{n,m}(t)\) is the available bandwidth of the V2I link at time $t$.

Propagation delay refers to the time required for a signal to propagate through the physical medium to the receiving node, which is related to the distance between nodes and the signal propagation speed. Given the short distance between nodes in VANET, the signal propagation delay is negligible \cite{feng2023wireless}. Therefore, the end-to-end delay from the source node \(v_s(t)\) to the destination node \(v_d(t)\) in VANET is given by:
\begin{equation}
	\begin{aligned}
		T_{sd}(t) &= \sum_{\substack{v_n(t) \in P_{sd}(t), \\ v_n(t) \neq v_d(t)}} T_{\text{queue},n}^V(t) + \sum_{r_m(t) \in P_{sd}(t)} T_{\text{queue},m}^I(t) \\
		&+ \sum_{\substack{(v_n(t),v_u(t)) \in \text{Link}(P_{sd}(t))}} T_{\text{trans},n,u}^V(t) \\
		&+ \sum_{\substack{(v_n(t),r_m(t)) \in \text{Link}(P_{sd}(t))}} T_{\text{trans},n,m}^I(t)
	\end{aligned}
\end{equation}
where \(\text{Link}(P_{sd}(t))\) denotes all active relay links included in the transmission path \(P_{sd}(t)\).

\noindent\textit{(2) Average Path Length}

The average path length in VANET is defined as the expectation of the shortest path hop count of key communication pairs (vehicle pairs that need cooperation despite exceeding the communication range of a single vehicle). It reflects the transmission efficiency of key multi-hop links, and its increase can effectively capture the congestion dynamics at the path level. Let $C(t)$ be the set of key communication pairs at time $t$, \(P_{sd}^*(t)\) be the optimal path from the source vehicle \(v_s(t)\) to the destination vehicle \(v_d(t)\) under the current topology \(\mathcal{G}(t)\), and \(L(P_{sd}^*(t))\)  be the minimum number of forwarding nodes or hop count of this path. Then the average path length is given by: 
\begin{equation}
	L_{\text{avg}}(t)= \frac{1}{|C(t)|}\sum_{(v_s(t),v_d(t)\in C(t)}L(P^*_{sd}(t))
\end{equation}
When \(L_{\text{avg}}(t)\) decreases, the hop counts or the number of intermediate nodes of information transmission paths for all key communication pairs are reduced overall.

\noindent\textit{(3) Network Throughput}

Throughput refers to the total volume of data packets successfully transmitted per unit time, which reflects the actual data transmission capacity of VANET \cite{RenRIS}. The total transmission volume per unit time of activated V2V links is given by:
\begin{equation}
	W_V(t) = \sum_{\substack{n=1,u=1 \\ n \neq u}}^{N(t)} x_{n,u}(t) \cdot B_{n,u}(t) \cdot \eta_V \cdot \mathbb{I}(d_{n,u}^V(t) \leq R_{\text{V2V}}(t))
\end{equation}
where \(\eta_V\) is the V2V link utilization coefficient, \(\mathbb{I}(\cdot)\) is the indicator function which takes the value of 1 if the inter-vehicle distance \(d_{n,u}^V(t) \leq\) the V2V communication radius \(R_{\text{V2V}}(t)\), and 0 otherwise. The total transmission volume per unit time of activated V2I links is given by:
\begin{equation}
	W_I(t) = \sum_{n=1}^{N(t)} \sum_{m=1}^M y_{n,m}(t) \cdot b_{n,m}(t) \cdot \eta_I \cdot \mathbb{I}(d_{n,m}^I(t) \leq R_{\text{V2I}}(t))
\end{equation}
where \(\eta_I\) is the V2I link utilization coefficient. In addition, this paper takes the packet successful delivery rate in path transmission into account. The average packet loss rate at time $t$ can be expressed as:
\begin{equation}
	\rho_{\text{loss}}(t) = p \cdot L_{\text{avg}}(t)
\end{equation}
where $p$ is the packet loss probability per hop. This formula is a simplified model established based on the packet loss characteristics of VANET multi-hop communication, corresponding to the core rule that a shorter path leads to less packet loss. In summary, the total network throughput of VANET can be expressed as:
\begin{equation}
	W_{total}(t)=[W_V(t)+W_I(t)]\cdot (1- \rho_{loss}(t))
\end{equation}

\noindent\textit{(4) Joint Optimization of Average Path Length and End-to-End Delay}

The dynamic topology and fluctuating communication demands in VANETs can easily lead to network congestion. To address this, a dynamic multi-objective mixed-integer optimization model is formulated. By dynamically adjusting the activation states of V2V/V2I links and their bandwidth allocation, the model simultaneously optimizes the average path length and end-to-end delay under various network constraints, thereby systematically mitigating congestion and improving communication performance. To jointly optimize these two metrics, their unit discrepancy must be eliminated. Dynamic normalization constants, $T_{\text{norm}}(t)$ for delay and $L_{\text{norm}}(t)$ for path length, are introduced for this purpose.

Regarding $T_{\text{norm}}(t)$, this work collects a set of end-to-end delay samples ${ T_1, T_2, ..., T_k }$ from the network controller over a period preceding time $t$. Anomalous samples are then filtered using the $3\sigma$ criterion. Finally, the value is updated via an exponentially weighted moving average formula that integrates real-time statistical values with historical baselines \cite{noor2021memory}:
\begin{equation}
	T_{\text{norm}}(t) = \beta \cdot T_{\text{max,real}}(t) + (1-\beta) \cdot T_{\text{norm,hist}}(t)
\end{equation}
where \(T_{\text{max,real}}(t)\) is the maximum value within the valid sample set at time $t$. $T_{\text{norm,hist}}(t)$ denotes the previously updated normalization constant at $t$, and $\beta$ is a dynamic weight that is adaptively adjusted by the real-time network fluctuation index \(\sigma_T(t)\).

Considering the strong correlation between path length and network topology connectivity, $L_{\text{norm}}(t)$ is defined by integrating topological connectivity with real-time network structure:
\begin{equation}
	L_{\text{norm}}(t) = \max\{L_{\text{max,real}}(t),\ \gamma \cdot Z(t)\}
\end{equation}
where $L_{\text{max,real}}(t)$ is the maximum path length among all service node pairs at time $t$, $Z(t)$ represents the network diameter at time $t$, and $\gamma$ is a safety coefficient introduced to prevent $L_{\text{norm}}(t)$ from falling below actual requirements due to local topological changes.

Furthermore, dynamic weighting coefficients \(\lambda_1(t)\) and \(\lambda_2(t)\) are introduced to adapt to the dynamic quality-of-service requirements of VANETs. These coefficients are updated in a stepwise manner to prevent gradient abruptness in the objective function \cite{amid2022step}:
\begin{equation}
	\lambda_1(t) = \lambda_1(t-1) + \Delta\lambda \cdot \text{sign}(Q_{\text{urgent}}(t) - Q_{\text{urgent,th}})
\end{equation}
where \(Q_{\text{urgent}}(t)\) denotes the proportion of urgent services at time $t$, \(Q_{\text{urgent,th}}\) is the predefined threshold for this proportion, and $\Delta\lambda$ represents the adjustment step size. These dynamic coefficients are constrained such that \(\lambda_1(t) + \lambda_2(t) = 1\). Consequently, the update rule for \(\lambda_2(t)\) can be directly derived from that of \(\lambda_1(t)\). The composite objective function for joint optimization is thus formulated as:
\begin{subequations}
	\label{deqn_ex15}
	\renewcommand{\theequation}{15\alph{equation}}
	\begin{align}
		&\underset{S(t)}{\min}\ \left[ \lambda _1(t) \cdot \frac{L_{\text{avg}}(t)}{L_{\text{norm}}(t)}+\lambda _2(t) \cdot \frac{\mathbb{E}[ T_{\text{sd}}(t)]}{T_{\text{norm}}(t)} \right] \label{deqn_ex15a} \\
		&\text{s.t.} \quad 
		\lambda _1(t) +\lambda _2(t) =1; \label{deqn_ex15b} \\
		&\quad\ \ d_{n,u}^{V}(t) \le R_{V2V},\ \forall n,u\in N; \label{deqn_ex15c} \\
		&\quad\ \ d_{n,m}^{I}(t) \le R_{V2I},\ \forall n\in N,\ \forall m\in M; \label{deqn_ex15d} \\
		&\quad\ \ \forall (v_n(t) ,v_u(t)) \in C(t), \exists k\geq k_{\min}, \label{deqn_ex15e} \\
		&\quad\ \ P_{sd,1}(t) ,...,P_{sd,k}(t) \subseteq G(t); \label{deqn_ex15f} \\
		&\quad\ \ \sum_{u=1}^N{x_{n,u}}\le \varDelta _{V2V},\ \forall n\in N ; \label{deqn_ex15g} \\
		&\quad\ \ \sum_{n=1}^N{y_{n,m}}\le \varDelta _{V2I},\ \forall m\in M; \label{deqn_ex15h} \\
		&\quad\ \ \sum_{n=1}^N{b_{n,m}}\le B_{RSU},\ \forall m\in M. \label{deqn_ex15i}
	\end{align}
\end{subequations}
where (15b) ensures a dynamically balanced joint optimization of the average path length and the end-to-end delay, preventing excessive bias towards a single metric. (15c) stipulates that activating a V2V link requires the actual distance between vehicles not to exceed the maximum communication range $R_{\text{V2V}}$. (15d) specifies that a V2I link can only be activated if the vehicle-to-RSU distance is within the maximum range \(R_{\text{V2I}}\). (15e) and (15f) are connectivity constraints, where critical communication pairs must maintain at least \(k_{\text{min}}\) available paths, ensuring topological connectivity and cooperative task reliability. (15g) limits the number of activated V2V links per vehicle node to its maximum indegree \(\Delta_{\text{V2V}}\), preventing buffer overload and high queuing delays. (15h) restricts the number of activated V2I links per RSU to its maximum indegree \(\Delta_{\text{V2I}}\) avoiding excessive load on the RSU. (15i) enforces that the total bandwidth allocated by an RSU to all connected vehicles does not exceed its maximum capacity \(B_{\text{RSU}}\).

\section{Hierarchical Optimization Based VANET Topology Dynamic Regulation}\label{3}

To achieve real-time, distributed solutions for the dynamic multi-objective optimization problem formulated in Section II, this paper proposes a VANET topology dynamic regulation scheme based on hierarchical optimization. This scheme decomposes the global topology optimization task into three hierarchical layers: local node feature extraction, node feature fusion, and global topology adjustment. This approach ensures optimization accuracy while significantly reducing communication and computational overhead, thereby accommodating the highly dynamic and distributed nature of VANETs.

\subsection{Local Node Feature Extraction}

The real-time state information of vehicle nodes and RSUs in VANET exhibits significant spatiotemporal dynamics. To support the training and decision-making of subsequent topology optimization models, it is necessary to extract feature information from each network node that comprehensively reflects its dynamic behavior, spatial location, and interaction requirements. These features serve as inputs to the optimization algorithm and directly influence the accuracy and effectiveness of the topology adjustment strategy. To this end, this paper constructs a node feature matrix $\mathbf{X}(t) = [\mathbf{x}_1, \mathbf{x}_2,..., \mathbf{x}_{N(t)+M}]^T$, where each row corresponds to the feature vector of one node.

The communication demand intensity between vehicle nodes is jointly influenced by factors such as spatial proximity and relative motion states. To quantify this relationship, this paper generates a symmetric inter-vehicle communication demand matrix $\mathbf{D}(t)$ of size $N(t) \times N(t)$ based on vehicle position and velocity information at time $t$. The off-diagonal element $D_{ij}(t)$ ($i \neq j$) is computed by jointly considering the Euclidean distance and velocity similarity between vehicles $v_i$ and $v_j$, while $D_{ii}(t) = 0$.

To avoid dynamic changes in feature dimension with the number of vehicles $N(t)$ , this paper employs an inbound demand aggregation method to extract fixed-dimensional demand features for each vehicle. Specifically, the arithmetic mean is calculated for the $n$-th column of the demand matrix $\mathbf{D}(t)$, which represents the communication demand from all other vehicles toward the target vehicle $v_n$, yielding a scalar value:
\begin{equation}
	d_n(t)=\frac{1}{N(t)}\sum_{i=1}^{N(t)}D_{in}(t)
\end{equation}
This value reflects the global demand intensity for vehicle $v_n$ as a communication destination. Based on the feature extraction results above, the feature vector for each node is constructed as a fixed-dimensional vector containing five key state information dimensions:
\begin{equation}\label{x}
	\mathbf{x}_n(t)=[p_n^x(t), p_n^y(t), s_n(t), \theta_n(t), d_n(t)]
\end{equation}
where $p_n^x(t), p_n^y(t)$ represent the node's coordinates in the two-dimensional plane. For vehicle nodes, this information is acquired in real time from onboard GPS units; for RSU nodes, it corresponds to their fixed deployment coordinates. $s_n(t)$ characterizes the node's instantaneous speed, with the speed of RSU nodes being constant at zero. $\theta_n(t)$ denotes the heading angle of vehicle nodes; RSU nodes have no directional information. $d_n(t)$ is the node communication demand intensity value calculated via Eq. (\ref{x}). This construction ensures that the feature vector length for each node remains constant at 5, regardless of changes in the network size $N(t)$, thereby satisfying the requirement for consistent input dimensionality in subsequent feature fusion and the optimization model.

\subsection{Node Feature Fusion}

In highly dynamic VANETs, the local perception of vehicle nodes is limited by their communication range, leading to insufficient spatiotemporal information about the network state. To support subsequent global optimization decisions for dynamic topology, it is essential to construct an enhanced feature representation that captures both local environment and global correlations through inter-node collaboration and information exchange. To this end, this paper proposes a hierarchical feature fusion mechanism based on dynamic neighborhood awareness. Through a limited number of rounds of local information exchange and adaptive weighted fusion, this mechanism enables the feature vector of each node to retain its core state while integrating environmental context from its neighborhood. This results in robust and accurate feature inputs for topology optimization.

\textit{(1) Fusion Framework.} Feature fusion is performed independently at each discrete time step $t$. For any node, whether a vehicle or an RSU (denoted as $v_n$ for convenience), the fusion process is defined by a mapping function $\mathcal{F}$. This function takes its current feature vector and the feature vectors of all its neighbor nodes as inputs, and outputs an updated feature vector:
\begin{equation}
	\mathbf{x}_n(t+1)=\mathcal{F}(\mathbf{x}_n(t), \{\mathbf{x}_u(t)| v_u \in \mathcal{N}_n(t)\})
\end{equation}
where $\mathcal{N}_n(t)$ denotes the set of all neighbor nodes within the communication range of node $v_n$ at time $t$. The mapping $\mathcal{F}$ is specifically composed of two sequential core modules: neighborhood feature aggregation and adaptive feature fusion.

\textit{(2) Neighborhood Feature Aggregation.} The objective of neighborhood feature aggregation is to combine the feature information from neighboring nodes, thereby providing environmental context for the central node. Considering the distance-dependent attenuation of wireless signal strength, this paper employs a distance-decay model to quantify the contribution of different neighbors to the fusion process. For a node $v_n$ and its neighbor $v_u$, their contribution weight $\omega_{n,u}$ is calculated as:
\begin{equation}\label{dd}
	\omega_{n,u}=\frac{\exp(-\lambda\cdot d_{n,u})}{\sum_{v_i\in\mathcal{N}_n(t)}\exp(-\lambda\cdot d_{n,i})}
\end{equation}
where $d_{n,u}$ is the Euclidean distance between the two nodes and $\lambda>0$ is the distance decay coefficient. This model ensures that closer neighbors have higher weights and thus a more significant impact on the feature update of the central node. Based on these weights, the neighborhood-aggregated feature vector $\tilde{\mathbf{x}}_n(t)$ for node $v_n$ is obtained by calculating the weighted average of the features from all its neighbor nodes:
\begin{equation}\label{nn}
	\tilde{\mathbf{x}}_n(t)=\sum_{v_u\in\mathcal{N}_n(t)}\omega_{n,u} \cdot \mathbf{x}_u(t)
\end{equation}

\textit{(3) Adaptive Feature Fusion.} After obtaining the neighborhood-aggregated features, a node needs to fuse them with its own features. Given the inherent differences in roles and state stability between vehicle nodes and RSU nodes within the network, a fusion strategy with fixed weights struggles to adapt to dynamic scenarios. Therefore, this paper introduces a type-dependent adaptive weight coefficient $\omega_n^{self}$ to perform a weighted fusion of the node's own features and the neighborhood-aggregated features:
\begin{equation}\label{xn}
	\mathbf{x}_n(t+1)=\omega_n^{self} \cdot \mathbf{x}_n(t)+(1-\omega_n^{self}) \cdot \tilde{\mathbf{x}}_n(t)
\end{equation}
This coefficient is dynamically set based on the node type. A higher value of $\omega_n^{self}$ is assigned to vehicle nodes to emphasize the primacy of their own high-mobility states. Conversely, a lower value is assigned to RSU nodes, reflecting their role as fixed infrastructure that focuses more on aggregating and integrating information from surrounding vehicles.

\textit{(4) Asynchronous Iterative Updates and Convergence.} The entire feature fusion process is executed through multiple iterative rounds, as detailed in Algorithm \ref{alg1}. The algorithm employs a distributed asynchronous update mechanism. In each iteration, every node independently and in parallel computes a new feature vector based solely on its locally perceived neighbor states. This eliminates the need for network-wide node synchronization or central controller coordination, thereby significantly reducing communication overhead and enhancing system responsiveness in dynamic networks. The iterative process incorporates dual termination conditions to ensure a balance between efficiency and accuracy:
\begin{enumerate}
	\item Maximum iteration count: Reaching the predefined maximum number of iterations $T_{max}$.
	\item Feature convergence criterion: The update variation of the feature vectors for all nodes falls below a predefined convergence threshold $\epsilon $; that is, the condition
	\begin{equation}
		\max_{v_n\in\mathcal{V}(t)}\left \| \mathbf{x}_n(t+1)-\mathbf{x}_n(t) \right \| <\epsilon
	\end{equation}
\end{enumerate}
The iteration terminates once either condition is met. The final output, the fused feature matrix $\mathbf{X}^*(t)$, serves as the critical input for the global topology adjustment model, providing a high-quality joint feature representation for link quality assessment and optimization decision-making.

\begin{algorithm}[t]
	\caption{Hierarchical Feature Fusion Algorithm}\label{alg1}
	\begin{algorithmic}[1]
		\STATE \textbf{Input:} Current network topology $\mathcal{G}(t)$, Initial node feature matrix $\mathbf{X}(t)$, Maximum iteration rounds $T_{max}$, convergence threshold $\epsilon$
		
		\STATE \textbf{Output:} The fused node feature matrix $\mathbf{X}^*(t)$.
		
		\WHILE{$ t< T_{max} $}
		\FOR {Each node in $\mathcal{G}(t)$ executes in parallel} 
		\STATE Obtain its set of neighbor nodes $\mathcal{N}_n(t)$.
		\STATE $\omega_{n,u}$ is calculated by Eq. \ref{dd}.
		\STATE $\tilde{\mathbf{x}}_n(t)$ is calculated by Eq. \ref{nn}.
		\STATE $\mathbf{x}_n(t+1)$ is updated by Eq. \ref{xn}.
		\ENDFOR
		\STATE The maximum feature change in this iteration is computed: $\Delta_{max}=\max_{v_n}\left \| \mathbf{x}_n(t+1)-\mathbf{x}_n(t) \right \|$.
		\IF{$\Delta_{max}<\epsilon$}
		\STATE break
		\ENDIF
		\ENDWHILE	
	\end{algorithmic}
\end{algorithm}

\subsection{Global Topology Adjustment}
As the third stage, the global topology adjustment aims to instantiate and solve the joint optimization problem in Eq. (15) based on the enhanced feature matrix output by feature fusion.

To meet the dual requirements for real-time performance and accuracy of the optimization algorithm in dynamic VANET scenarios, a dual-mode adaptive solving strategy is designed. This strategy dynamically selects between exact and heuristic solving modes based on the current network state complexity. A complexity criterion, $Q(t)$, is defined as the basis for mode switching:
\begin{equation}
	Q(t)=\xi \cdot N(t)+\zeta \cdot \rho (t)
\end{equation}
where $\rho(t)$ denotes the network link density, and $\xi$ and $\zeta$ are weighting coefficients used to adjust the contribution of network scale and connection density to computational complexity, respectively. When $Q(t)$ falls below a preset threshold $Q_0$, it indicates a manageable problem scale, and the exact solving mode is activated. In this mode, the optimization problem is formulated as a Mixed-Integer Linear Programming model and solved to global optimality using a commercial solver, provided sufficient computational resources are available. Otherwise, the strategy switches to the heuristic solving mode to improve response speed. This mode employs approximation strategies to rapidly generate high-quality feasible solutions, significantly reducing computation time with acceptable accuracy trade-offs.

To balance the performance gain from topology updates against the risk of network oscillation due to frequent adjustments, a dual mechanism, comprising a performance improvement threshold and a topology validity verification, is introduced. First, the joint improvement rate $\delta(t)$ brought by a candidate optimization strategy is calculated. Aligned with the weights of the objective optimization function, $\delta(t)$ ensures that the decision direction remains consistent with the optimization goal. It is defined as:
\begin{equation}
	\delta(t)=\lambda_1(t)\cdot \frac{L_{\text{avg}}-L_{\text{avg}}^*(t)}{L_{\text{avg}}\cdot L_{\text{norm}}(t)} + \lambda_2(t)\cdot \frac{\mathbb{E}[T_{sd}(t)]-\mathbb{E}[T_{sd}^*(t)]}{\mathbb{E}[T_{sd}(t)] \cdot T_{\text{norm}}(t)}
\end{equation}
where $L_{\text{avg}}^(t)$ and $\mathbb{E}[T_{sd}^(t)]$ denote the optimized average path length and end-to-end delay, respectively.

Second, a three-step validity verification is performed on the candidate topology:
(1) Verify whether the candidate topology satisfies all preset model constraints.
(2) Predict the expected lifetime of each candidate link based on vehicle kinematic states. Temporary links with a lifetime shorter than two control cycles are eliminated to ensure short-term topological stability.
(3) If the candidate topology fails to satisfy the constraints, invalid links are frozen, and new links are supplemented from the candidate link pool that meets the distance constraints. This avoids the latency associated with a complete re-solving process.

A global topology update is triggered only when $\delta(t)$ exceeds a preset threshold $\delta_0$ and the candidate topology passes the validity verification, thereby mitigating the risk of network oscillation. The overall workflow of the global topology adjustment is outlined in Algorithm \ref{alg2}.

\begin{algorithm}[t]
	\caption{Global Topology Adjustment Algorithm}\label{alg2}
	\begin{algorithmic}[1]
		\STATE \textbf{Input:} Integrated node feature matrix $\mathbf{X}^*(t)$, current network topology $G(t)$, maximum simulation steps $T_{\max}$.
		
		\STATE \textbf{Output:} Optimized network topology $G^*(t)$.
		
		\WHILE{$ t< T_{max} $}
		\STATE Extract key information from $\mathbf{X}^*(t)$ to identify critical communication pairs.
		\STATE Compute $L_{\text{avg}}(t)$ and $\mathbb{E}[T_{sd}(t)]$ by the current $G(t)$.
		\STATE Calculate current network complexity criterion $Q(t)$.
		\IF{$Q(t) < Q_0$}
		\STATE Activate the exact solving mode, obtaining the candidate strategy $S^*(t)$.
		\ELSE
		\STATE  Activate the heuristic solving mode to rapidly generate the candidate strategy $S^*(t)$.
		\ENDIF
		\STATE Construct the candidate topology $G^*(t)$ based on the candidate strategy $S^*(t)$, and predict its performance $L_{\text{avg}}^*(t)$ and $\mathbb{E}[T_{sd}^*(t)]$.
		\STATE Compute the joint improvement rate $\delta(t)$.
		\STATE Perform topology validity verification on the candidate topology $G^*(t)$.
		\IF{$\delta(t) > \delta_0$ and verification passed }
		\STATE Trigger the topology update: $G(t+1) \leftarrow G^*(t)$.
		\ELSE
		\STATE Perform local correction to generate a revised topology $G^\prime (t)$, and re-verify it.
		\STATE If passed, update the topology; otherwise, maintain the original topology: $G(t+1) \leftarrow G(t)$.
		\ENDIF
		\ENDWHILE	
	\end{algorithmic}
\end{algorithm}

\begin{figure}[t]
	\centerline{\includegraphics[width=3.3in,keepaspectratio]{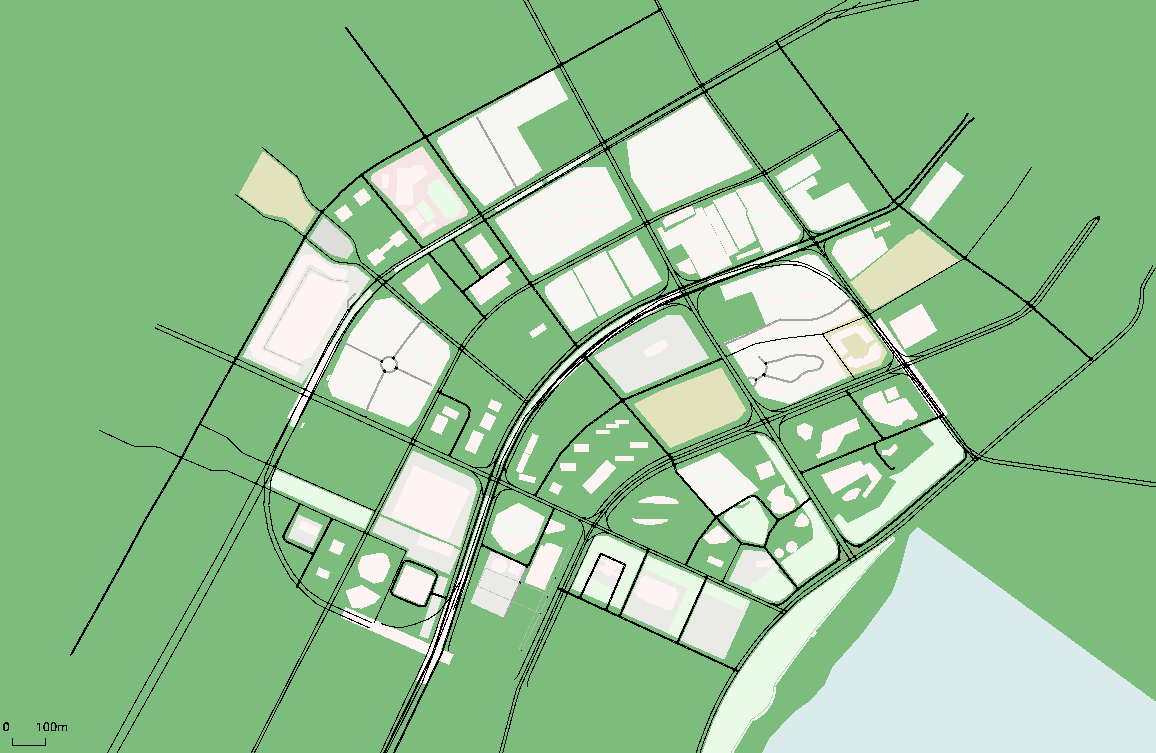}}
	\caption{Road Network of a Real-World Urban Traffic Scenario.}
	\label{fig1}
\end{figure}

\section{Simulation and Result Analysis}\label{4}
This section validates the effectiveness of the proposed hierarchical optimization-based VANET topology dynamic control scheme through systematic testing in a realistic urban traffic scenario constructed on the SUMO simulation platform. As shown in Fig. \ref{fig1}, the simulation uses the actual road network between Weidong Station and Honggu Middle Road Station in Nanchang City, Jiangxi Province, China. The OSMWebWizard tool reconstructs a dynamic traffic environment that includes multi-lane roads, intersections, and traffic signals. By importing a standard vehicle motion model and setting a vehicle generation rate consistent with urban traffic flow, the simulation reproduces vehicle density and movement patterns during peak hours. The SUMO platform collects dynamic parameters, including vehicle coordinates and speed profiles, to generate a standardized vehicle motion topology file. Based on this setup, we perform a comparative analysis with the following three classic topology optimization algorithms:
\begin{itemize}
	\item Greedy Algorithm: Based on the principle of local optimality, it prioritizes establishing the shortest-distance or highest-quality V2V/V2I links at each time step \cite{Greedy}.
	\item Shortest Path Algorithm: Aiming to minimize hop count, it constructs the topology using graph-theoretic shortest-path algorithms (e.g., Dijkstra).
	\item Motif-based Algorithm: By identifying frequently occurring vehicle movement patterns in the network, it derives link establishment rules from historical patterns, exhibiting certain scenario adaptability.
\end{itemize}

The core simulation parameters, configured according to typical urban VANET communication standards and device specifications, are presented in Table \ref{tab1}.
\begin{table}[t]
	\centering
	\caption{Simulation Parameter Configuration}
	\label{tab1}
	\renewcommand{\arraystretch}{1.2}
	\begin{tabular}{|c|c|}
		\hline
		\textbf{Parameter} & \textbf{Value} \\
		\hline
		V2V maximum communication range $R_{V2V}$ & 300 m \\
		\hline
		RSU coverage range $R_{V2I}$ & 500 m \\
		\hline
		Maximum vehicle node in-degree $\Delta_{V2V}$ & 5 \\
		\hline
		Maximum RSU node in-degree $\Delta_{V2I}$ & 10 \\
		\hline
		RSU maximum bandwidth capacity $B_{RSU}$ & 100 Mbps \\
		\hline
		Speed weight coefficient $\alpha$ & 0.7 \\
		\hline
		Trajectory overlap threshold $R_{th}$ & 0.7 \\
		\hline
		Per-hop packet loss rate $p$ & 0.03 \\
		\hline
	\end{tabular}
\end{table}

\begin{figure}[t]
	\centerline{\includegraphics[width=3.3in,keepaspectratio]{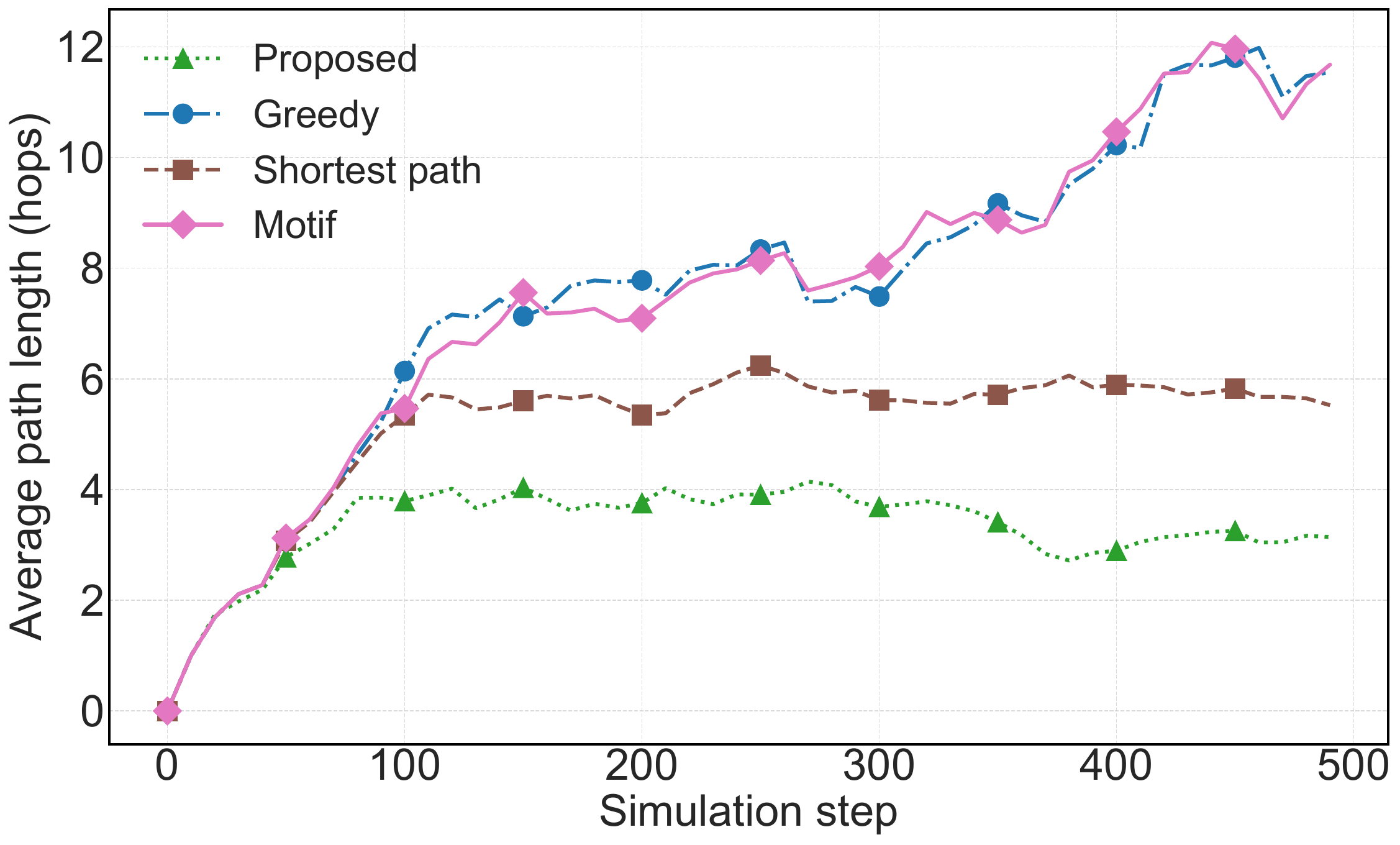}}
	\caption{Comparison of Average Path Length Across Different Algorithms.}
	\label{fig2}
\end{figure}

Fig. \ref{fig2} shows the evolution of the average path length over 500 simulation steps for the four algorithms. It can be observed that the proposed scheme converges rapidly and stabilizes at approximately 4 hops. In contrast, the greedy algorithm, which focuses only on local optimality, tends to form lengthy paths, resulting in fluctuations between 8 and 12 hops during the middle to late stages of the simulation. While the shortest-path algorithm aims to minimize hop count, its failure to account for link dynamics leads to an actual path length between 6 and 9 hops. The motif-based algorithm, constrained by fixed pattern rules, exhibits a path length exceeding 10 hops in dynamic scenarios. In comparison, the proposed scheme achieves effective and stable minimization of the average path length in dynamic environments by rapidly perceiving neighborhood changes through local feature fusion and performing cross-region coordination via global optimization.

\begin{figure}[t]
	\centerline{\includegraphics[width=3.3in,keepaspectratio]{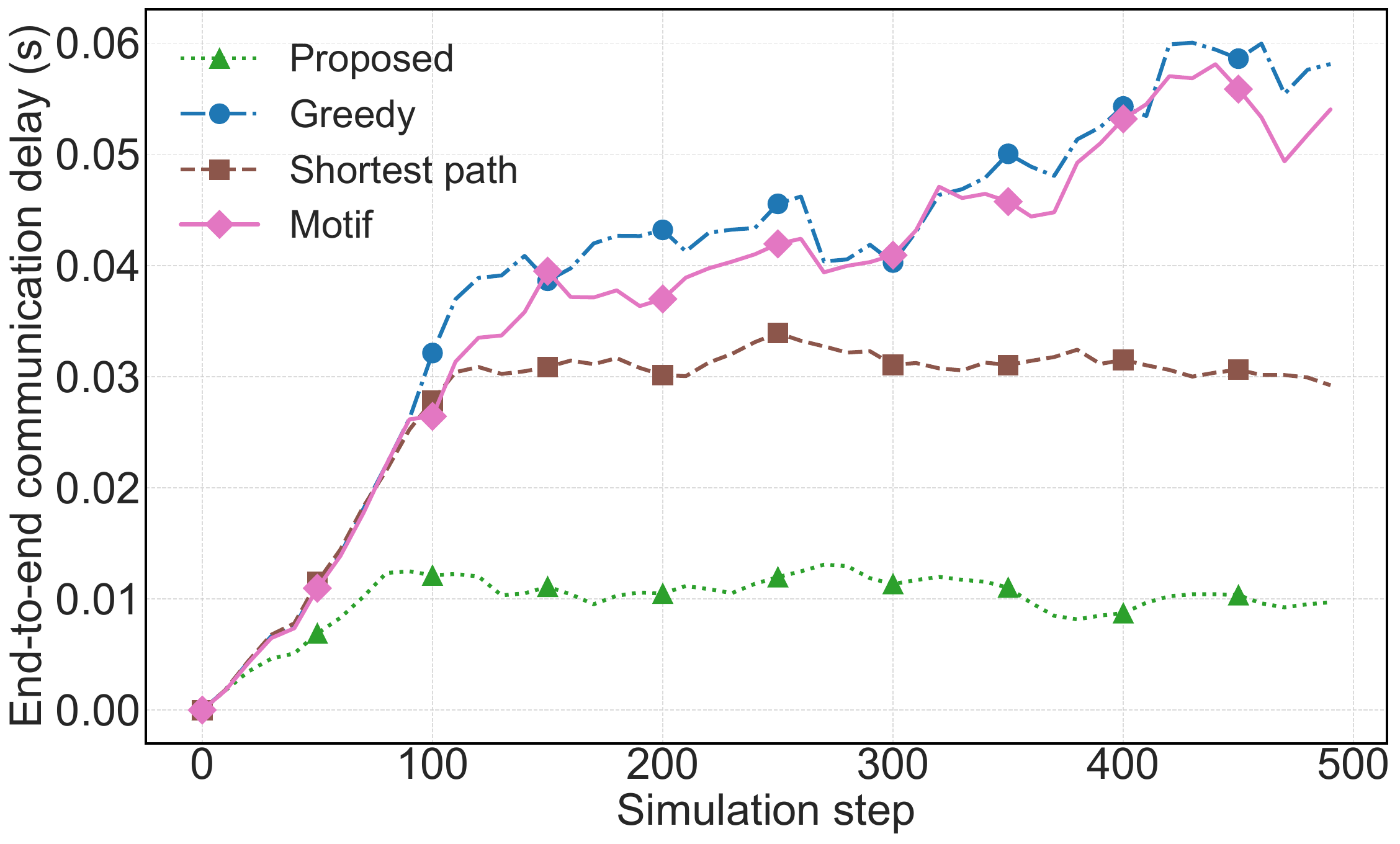}}
	\caption{Comparison of End-to-End Delay Under Different Algorithms.}
	\label{fig3}
\end{figure}

Fig. \ref{fig3} compares the variation of end-to-end delay over time for the four algorithms. The proposed scheme consistently maintains a low delay, stabilizing around 0.01 seconds. This performance is attributed to its local feature filtering mechanism, which can predict link congestion risks and perform proactive traffic steering. Combined with millisecond-level global topology coordination, it ensures network-wide load balancing, thereby effectively suppressing both the increase and fluctuation of end-to-end delay. In contrast, the delays of the three baseline algorithms exhibit an upward trend as the simulation progresses, with those of the greedy and motif-based algorithms accelerating notably after step 300 as network load intensifies. This degradation stems from their tendency to cause traffic imbalance and local congestion in dynamic environments: the greedy algorithm directs traffic toward links with momentarily optimal but potentially soon-to-break quality; the shortest-path algorithm may concentrate excessive traffic on a few critical paths; and the motif-based algorithm lacks congestion avoidance capability due to its rule-based link selection.

\begin{figure}[t]
	\centerline{\includegraphics[width=3.35in,keepaspectratio]{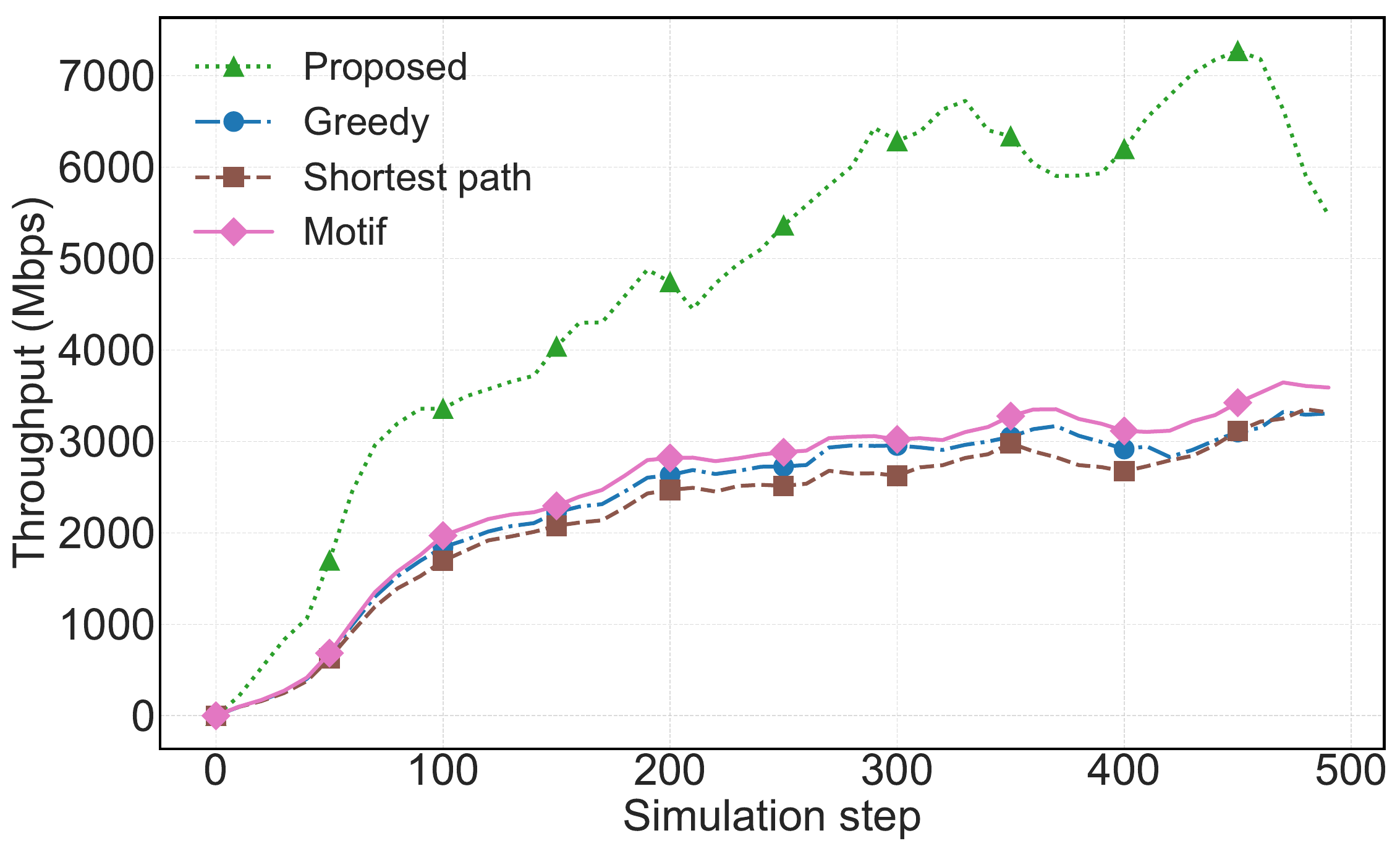}}
	\caption{Comparison of Network Throughput Under Different Algorithms.}
	\label{fig4}
\end{figure}

Fig. \ref{fig4} shows the evolution of network throughput over 500 simulation steps for the four algorithms. The results demonstrate that the proposed scheme exhibits a clear absolute advantage in network throughput throughout the simulation. In contrast, the throughput of the other algorithms remains at relatively low levels, with their growth curves tending to flatten. This superior performance stems from the proposed scheme's ability to enable nodes to perform intelligent forwarding based on local link quality and congestion status, thereby preventing data packets from flooding potential bottleneck links and avoiding the throughput ceiling caused by the overload of a few critical paths. Furthermore, global adjustment dynamically constructs multiple efficient short-path routes, reducing dependence on a limited number of key paths, decreasing transmission hops, and freeing up network resources, which collectively leads to a significant throughput improvement. While the comparative algorithms show some improvement, their performance ceiling is evident. This limitation arises because traditional algorithms, while optimizing local or single metrics (e.g., hop count), often neglect overall resource utilization efficiency.

\begin{figure}[t]
	\centerline{\includegraphics[width=3.3in,keepaspectratio]{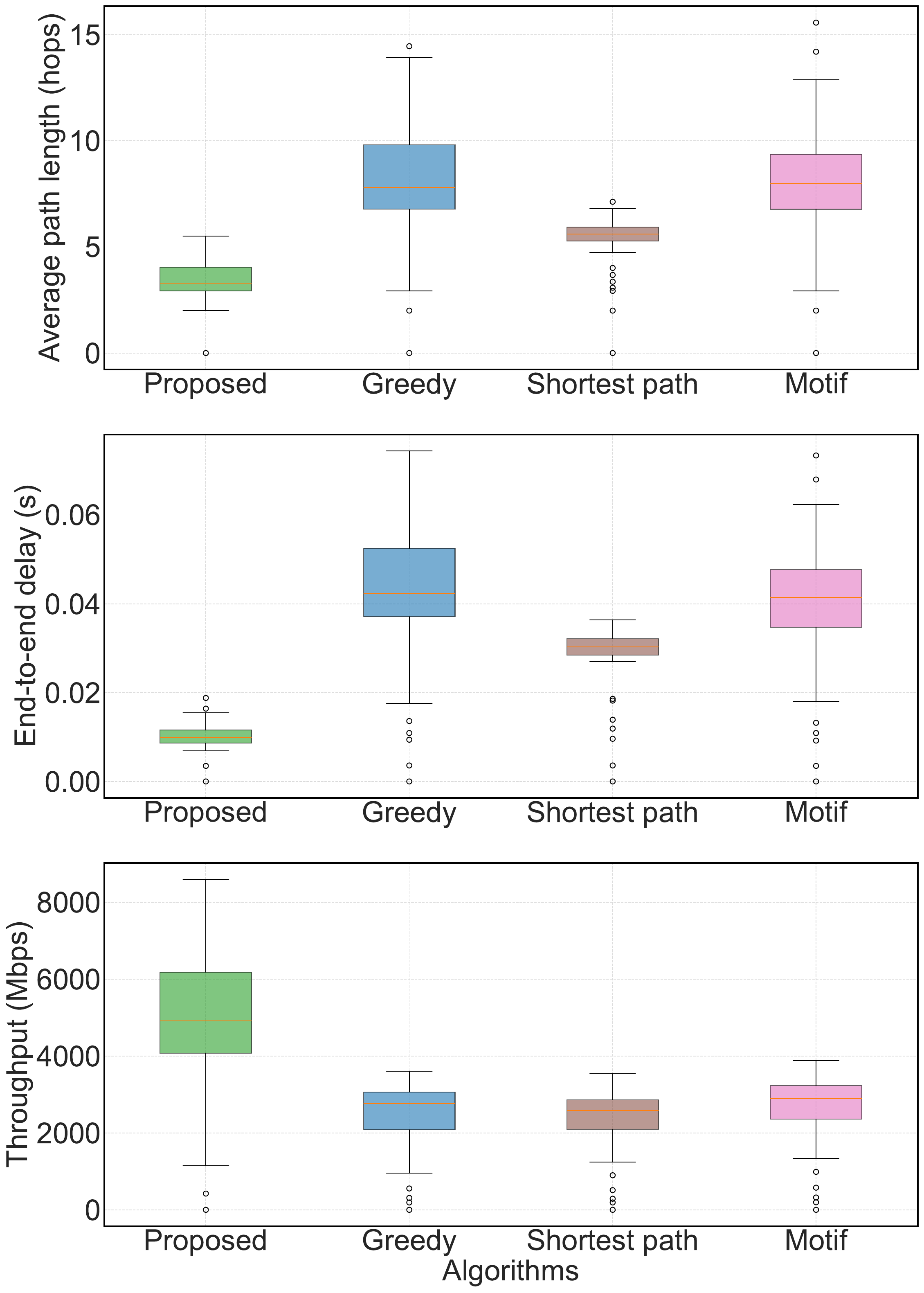}}
	\caption{Boxplots of the Three Performance Metrics Across Different Algorithms.}
	\label{fig5}
\end{figure}

Fig. \ref{fig5} presents the statistical distribution of each performance metric throughout the entire simulation via boxplots. For end-to-end delay, the proposed scheme exhibits the shortest box and the lowest median, demonstrating the best stability. Regarding average path length, it achieves the lowest median, reflecting superior average performance. Although its box length is slightly greater than that of the shortest-path algorithm, it remains significantly better than both the greedy and motif-based algorithms. In terms of network throughput, the box for the proposed scheme is positioned substantially higher than those of the other algorithms, confirming a clear capacity advantage. These results collectively validate that the proposed scheme, through its local rapid perception and global cooperative optimization, achieves a balanced improvement in low latency, short paths, and high throughput in dynamic VANETs, outperforming traditional methods in both overall performance and stability.

\begin{figure}[t]
	\centerline{\includegraphics[width=3.2in,keepaspectratio]{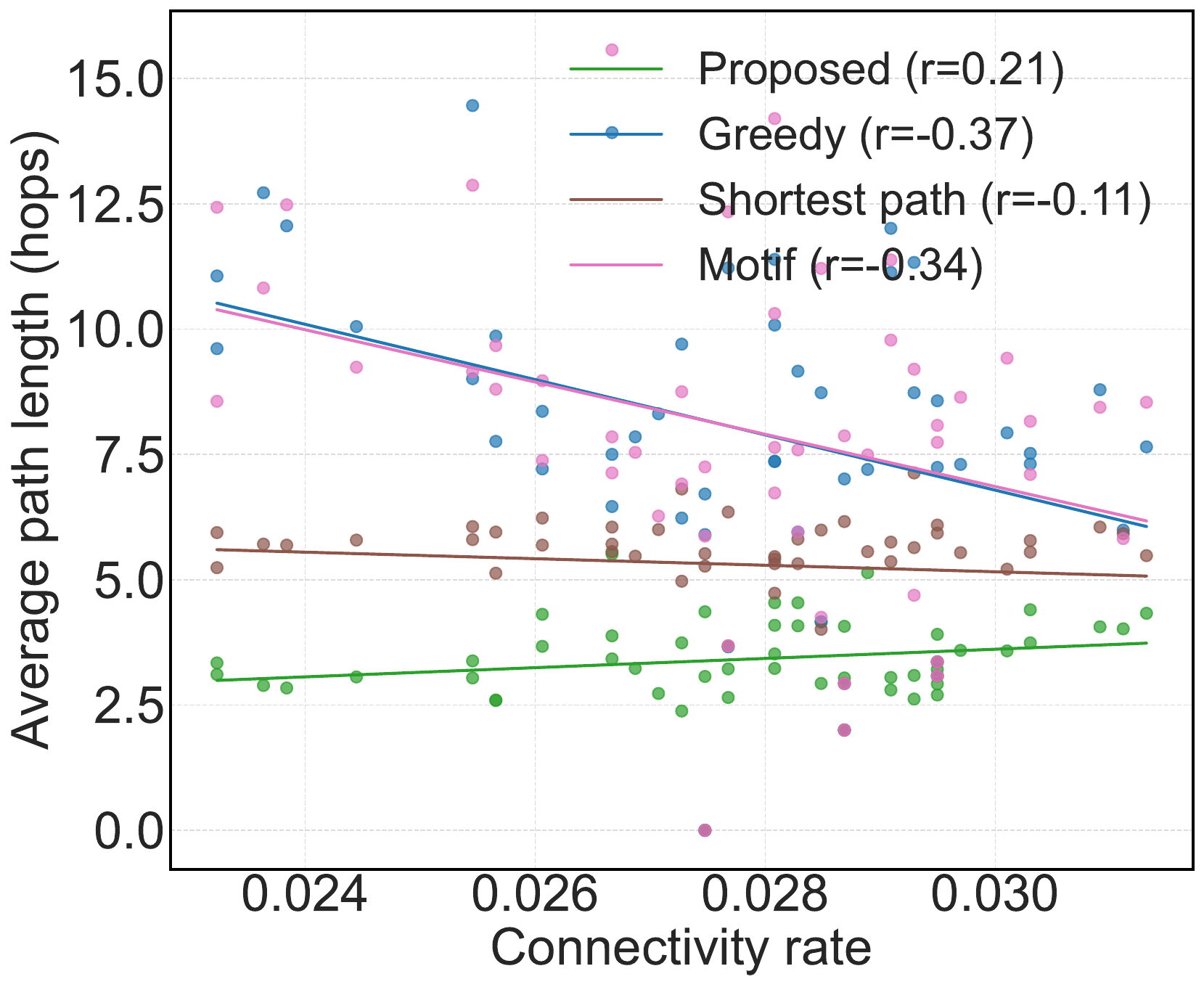}}
	\caption{Average Path Length versus Connectivity Rate: Scatter Plot and Correlation Analysis.}
	\label{fig6}
\end{figure}

An inherent trade-off exists between network connectivity rate and average path length. Fig. \ref{fig6} plots the scatter diagram of average path length versus connectivity rate for each algorithm, along with linear trend lines and correlation coefficients. The scatter points for the proposed scheme are the most concentrated and show a slight positive correlation. This indicates that at high connectivity levels, the algorithm tends to introduce a small number of V2I relays to secure higher bandwidth. Although this slightly increases the hop count, the path length remains stable below 5 hops, which is far lower than the baselines. In contrast, both the greedy and motif-based algorithms exhibit a strong negative correlation. Their path lengths can reach up to 15 hops in sparse networks and are passively compressed to 7–8 hops as topology density increases, verifying their extremely low efficiency under low connectivity. The shortest-path algorithm shows the weakest correlation, reflecting its limited sensitivity to network connectivity due to its fixed shortest-hop strategy. This correlation analysis provides evidence from another perspective that the proposed scheme can adapt to varying network states by dynamically adjusting its optimization strategy, thereby consistently maintaining overall optimal communication performance.

\section{Conclusion}\label{5}
To address the issues of increased average path length, elevated end-to-end delay, and reduced throughput due to highly dynamic topologies in VANETs, A multi-objective optimization problem model integrating these three metrics was formulated. A hierarchical optimization framework was designed, where the local layer rapidly responded to changes through node feature extraction and dynamic neighborhood-aware feature fusion, while the global layer employed a dual-mode adaptive solving strategy to balance accuracy and real-time performance. Performance improvement thresholds and a topology validity verification mechanism were introduced to mitigate network oscillation risks. Simulation results on a real urban road network using the SUMO platform showed that the proposed scheme stabilized the average path length at approximately 4 hops, maintained end-to-end delay around 0.01 s, and achieved significantly higher network throughput compared to Greedy, Shortest-path, and Motif algorithms.

\bibliographystyle{ieeetr} 
\bibliography{MyRefs} 
~~~\\
~~~\\

\end{document}